# InGaAsP annular Bragg lasers: theory, applications and modal properties

Jacob Scheuer, William M. J. Green, Guy A. DeRose and Amnon Yariv

*Abstract*— A novel class of circular resonators, based on a radial defect surrounded by Bragg reflectors, is studied in detail. Simple rules for the design and analysis of such structures are derived using a transfer matrix formalism. Unlike conventional ring resonators, annular Bragg resonators (ABR) are not limited by the total internal reflection condition, and can exhibit both large free spectral ranges and low bend losses. The Bragg reflection mechanism enables the confinement of light within a defect consisting of a low refractive index medium (such as air). Strong atom-photon interaction can be achieved in such a structure, making it a promising candidate for sensing and cavity QED applications. For sensing applications, we show that the ABR structure can possess significantly higher sensitivity when compared to a conventional ring resonator sensor. Lasing action and low threshold levels are demonstrated in ABR lasers at telecommunication wavelengths under pulsed optical pumping at room temperatures. The impact of the intensity and dimensions of the pump-spot on the emitted spectrum is studied in detail.

## I. Introduction

THE past few years have witnessed a significant increase in research involving circular optical resonators. Resonators are key elements for various applications such as optical communication systems [1-6] and biochemical sensing [7,8], as well as for basic research such as single molecule spectroscopy [9] and cavity quantum electrodynamics (QED) [10,11] with possible applications in quantum information technologies.

For many of these applications, resonators that exhibit low losses (high $Q$-factor) are required. Particularly for sensing applications and for cavity QED, it is also often desired that the resonators have a small modal volume. These characteristics are required in order to attain strong atom-photon interaction and to probe as small a modal-volume as possible. For telecommunication applications, a large free spectral range (FSR) is often desired, which requires the resonators to be of small dimensions.

Circular resonators based on total internal reflection (TIR) that exhibit the combination of both high $Q$-factor and small dimensions are difficult to realize, because these attributes are mutually contradicting. To have large FSR, a small circumference and bending radius are required. Under such conditions, the efficiency of the TIR confinement mechanism is significantly reduced, leading to larger power radiation and lower $Q$-factors due to bending losses [12].

Photonic crystal (PC) cavities have been extensively studied for high $Q$ cavity applications. PC cavities with Q-factors approaching $\sim 10^6$ were predicted numerically [13], and $Q$s of up to $4.5 \times 10^4$ were demonstrated experimentally [14]. Nevertheless, these resonators consist primarily of a defect (either point or line), which does not necessarily support a whispering-gallery-mode-like (WGM) solution and are, therefore, difficult to couple to conventional waveguides. Hexagonal (non circular) cavities incorporating $120°$ abrupt bends, which can be easily coupled to PC line defect waveguides, have also been proposed and studied [15]. However, abrupt bends have been shown to support localized bound states [11] which might lase in addition to the desired WGM.

Recently, we proposed utilizing Bragg reflection instead of TIR as the radial confinement mechanism [16]. This concept is illustrated in Fig 1: A circumferentially guiding defect is located within a medium which consists of annular Bragg layers. Resonators of this class, known as annular Bragg resonators (ABR), are designed to support azimuthally propagating modes, with energy concentrated within the defect region by radial Bragg reflection. Unlike conventional resonators, the reflectivity of the Bragg mirrors can be increased simply by adding more layers. As a result, the radius of the defect, and therefore the modal volume, can be reduced significantly without increasing the bending losses.

Disk and ring resonators based on distributed Bragg reflection have been analyzed previously for both laser and passive resonator applications, employing various techniques such as conformal mapping, a coupled-mode approach and field transfer-matrices [17-24].

In this paper we explore theoretically and experimentally the properties of ABRs. In section II we briefly review the theoretical framework and the rules for the design and analysis of the resonators. In section III we study the advantages of the ABR structure for various applications, and in section IV we detail the fabrication of ABRs in semiconductor materials. In section V we present experimental results on ABR lasers, discussing our results in

The authors are with the Departments of Applied Physics and Electrical Engineering, MC 128-95, California Institute of Technology, Pasadena, CA 91125, USA (phone: 626-395-4413; fax: 626-405-0928; e-mail: koby@caltech.edu).

section VI.

## II. DESIGN AND ANALYSIS

### A. Theoretical Framework

We consider an azimuthally symmetric structure as illustrated in Fig. 1. The guiding defect, which consists of a material with refractive index $n_{def}$, is surrounded by distributed Bragg reflectors on both sides where the reflector layers are of refractive indices $n_1$ and $n_2$. All the electromagnetic field components can be expressed by the $z$ components of the electrical and magnetic fields [25] which satisfy the Helmholtz equations:

$$\left[ \frac{1}{\rho} \frac{\partial}{\partial \rho} \left( \rho \frac{\partial}{\partial \rho} \right) + \frac{1}{\rho^2} \frac{\partial^2}{\partial \theta^2} + k_0^2 n^2(\rho) + \frac{\partial^2}{\partial z^2} \right] \begin{bmatrix} E_z \\ H_z \end{bmatrix} = 0, \quad (1)$$

where $\rho$, $z$ and $\theta$ are the radial, axial and azimuthal coordinates respectively and $k_0$ is the wavenumber in vacuum. The refractive index $n(\rho)$ equals either $n_1$, $n_2$ or $n_{def}$ depending on the radius $\rho$. Assuming the dependence of the fields on the coordinates can be separated, the radial part of the fields (either $H_z$ or $E_z$), $R_{H,E}$, must satisfy the Bessel equation:

$$\rho^2 \frac{\partial^2 R_{H,E}}{\partial \rho^2} + \rho \frac{\partial R_{H,E}}{\partial \rho} + \left[ (k^2(\rho) - \beta^2)\rho^2 - m^2 \right] R_{H,E} = 0, \quad (2)$$

where $k(\rho) = k_0 \cdot n(\rho)$, $m$ is an integer and $\beta$ is the $z$ component of the wave-vector. The solutions of (2) are a superposition of the $m^{th}$ order Bessel function of the first and second kind:

$$E_z = \left[ A \cdot J_m \left( \sqrt{k_j^2 - \beta^2} \rho \right) + B \cdot Y_m \left( \sqrt{k_j^2 - \beta^2} \rho \right) \right]$$
$$\times \cos(\beta \cdot z + \varphi) \cdot \exp(im\theta) \quad (3)$$
$$H_z = \left[ C \cdot J_m \left( \sqrt{k_j^2 - \beta^2} \rho \right) + D \cdot Y_m \left( \sqrt{k_j^2 - \beta^2} \rho \right) \right]$$
$$\times \sin(\beta \cdot z + \varphi) \cdot \exp(im\theta)$$

where $A$, $B$, $C$, and $D$ are independent coefficients, $k_j$ is the material wavenumber in the $j^{th}$ layer. The other four components of the electric and magnetic fields can be readily derived from (3).

The parallel component of the fields – $E_z$, $H_z$, $E_\theta$, $H_\theta$ must be continuous at the interfaces separating successive layers. This requirement can be written in from of a transfer matrix, connecting the amplitude vector $[A\ B\ C\ D]$ in the $j^{th}$ and $j+1$ layers:

$$\begin{pmatrix} A_{j+1} \\ B_{j+1} \\ C_{j+1} \\ D_{j+1} \end{pmatrix} = \tilde{M}_{j+1}^{-1}(\rho_{j+1}) \cdot \tilde{M}_j(\rho_j) \cdot \begin{pmatrix} A_j \\ B_j \\ C_j \\ D_j \end{pmatrix} \quad (4)$$

and $\tilde{M}_j$ is given by:

$$\tilde{M}_j = \begin{pmatrix} J(\gamma_j \rho) & Y(\gamma_j \rho) & 0 & 0 \\ \frac{n_j^2}{\gamma_j} J'(\gamma_j \rho) & \frac{n_j^2}{\gamma_j} Y'(\gamma_j \rho) & -\frac{m\beta}{\rho \omega \varepsilon_0 \gamma_j^2} J(\gamma_j \rho) & -\frac{m\beta}{\rho \omega \varepsilon_0 \gamma_j^2} Y(\gamma_j \rho) \\ 0 & 0 & J(\gamma_j \rho) & Y(\gamma_j \rho) \\ -\frac{m\beta}{\rho \omega \mu \gamma_j^2} J(\gamma_j \rho) & -\frac{m\beta}{\rho \omega \mu \gamma_j^2} Y(\gamma_j \rho) & \frac{1}{\gamma_j} J'(\gamma_j \rho) & \frac{1}{\gamma_j} Y'(\gamma_j \rho) \end{pmatrix}, \quad (5)$$

where $\varepsilon$ and $\mu$ are the dielectric and magnetic susceptibilities,

$\omega$ is the optical angular frequency, $\gamma_j = \sqrt{k_j^2 - \beta^2}$, and the primes indicate derivative with respect to the function argument.

In the limit of strong vertical confinement (i.e., $\beta << k_j$) it is possible to separate the modal field solutions into two distinct polarizations: TE consisting of $H_z$, $E_\rho$ and $E_\theta$ and TM consisting of $E_z$, $H_\rho$ and $H_\theta$. Unlike [16] we adopt the polarization convention of planar optics.

In the above mentioned limit, each polarization component can be described by two coefficients in each layer: $A_j$ and $B_j$ for TM and $C_j$ and $D_j$ for TE. For each polarization, the boundary conditions at the interfaces between successive layers can be represented similarly to (4) using simplified 2X2 matrices:

$$\tilde{M}_j^{TM} = \begin{pmatrix} J(\gamma_j \rho) & Y(\gamma_j \rho) \\ \frac{n_j^2}{\gamma_j} J'(\gamma_j \rho) & \frac{n_j^2}{\gamma_j} Y'(\gamma_j \rho) \end{pmatrix} \quad \tilde{M}_j^{TE} = \begin{pmatrix} J(\gamma_j \rho) & Y(\gamma_j \rho) \\ \frac{1}{\gamma_j} J'(\gamma_j \rho) & \frac{1}{\gamma_j} Y'(\gamma_j \rho) \end{pmatrix} \quad (6)$$

Using relation (4) and the matrices (6), the field components can be "propagated" from the inner layers to the external layers. We use the finiteness of the field at $\rho=0$ so that $B_1 = D_1 = 0$. The second boundary condition is no inward propagating field beyond the last layer, so that $B_{N+1} = -iA_{N+1}$ for TM and $C_{N+1} = -iD_{N+1}$ for TE, where $N$ is the number of layers.

### B. Design Rules

The transfer matrix formalism enables us to find the modal field distribution in the case of an arbitrary arrangement of annular concentric dielectric rings. However, we are especially interested in structures that can lead to a concentration of the modal energy near a predetermined radial distance, i.e. within the defect.

It has been shown that the best strategy to attain an exponential decrease (or increase) in the field intensity in the grating region is to position the interfaces of the layers at the zeros and extrema of the $z$ component of the field [26]. The index profile and the field are calculated simultaneously, using the index to find the field and the field to determine the position of the interfaces.

It should be noted that the resulting optimal index profile exhibits an inherent resemblance to the conventional (Cartesian) Bragg reflector. The optimal Cartesian Bragg reflector can be designed in a similar way, leading to layers that are quarter-wavelength thick [27]. Here the resulting layers are also "quarter-wavelength" thick but in the sense of the quasi-periodicity of the $m^{th}$ order Bessel function [28]. The defect (again, as in the Cartesian case) should be "half-wavelength" wide, i.e. its interfaces should be located at successive zeros of the field.

In order to attain a transverse field profile which is confined within the defect, the profile must exponentially increase for $\rho < \rho_{def}$, and exponentially decrease for $\rho > \rho_{def}$. This requirement determines which index-interfaces (low→high or high→low) should be positioned at zeros of the field and which at the extrema of the field. The constraints on the index profile are similar to the Cartesian case and differ for the TE and TM polarizations [26]. For the TE polarization, the interfaces for increasing (decreasing) field should be at the zeros (extrema) of $H_z$ if $n(\rho^-) > n(\rho^+)$ at the interface and at the extrema (zeros) of $H_z$ if $n(\rho^-) < n(\rho^+)$ at the interface. For the TM polarization the interfaces for increasing (decreasing) field should be at the extrema (zeros) of $E_z$ if $n(\rho^-) > n(\rho^+)$ at the interface and at the zeros (extrema) of $E_z$ if $n(\rho^-) < n(\rho^+)$ at the interface. The interfaces of the defect must be located at zeros of $H_z$ for TE and of $E_z$ for TM.

### C. Mode Profile

Figure 2 depicts the refractive index (Fig. 2(a)) and the TE modal field (Fig. 2(b)) profiles of an ABR designed for a 0.55µm thick InGaAsP layer suspended in air. The device is designed to have a mode with an angular propagation coefficient of $m$=7 at $\lambda_{res}$=0.852µm. The effective index approximation in the vertical dimension is used to reduce the 3D problem to a 2D equivalent one. As can be seen in the figure, the field is primarily confined in the defect and it decays while oscillating in the Bragg reflectors.

To verify the validity of the effective index approximation we simulate the device using an angular finite-difference-time-domain (FDTD) scheme which takes advantage of the azimuthal symmetry to reduce the computational domain [29]. For a given $m$, the method can be used to find the resonance wavelength, the quality factor and the modal field profile of any circular device.

Figure 3 depicts a comparison between the field profiles calculated by the FDTD simulation and by the 2D transfer matrix formalism. There is good agreement between the two approaches. The resonance wavelength found by the FDTD simulations is 0.85µm and the ratio between $H_z$ and $E_z$ is 25dB, indicating that the modal field is primarily TE polarized.

Due to the different radial confinement mechanism, the characteristics of the ABR modal field differ significantly from those of the mode of a conventional resonator. First, the radial position of the maximal intensity of the field can be predetermined regardless of wavelength and material system. Second, low angular propagation coefficients ($m$) and tight bending radius can be realized because there is no need to satisfy a TIR condition. Finally, the field can be confined in a *lower* refractive index layer, giving rise to a larger FSR or enhanced sensitivity for sensing application (see section III).

## III. APPLICATIONS

As mentioned previously, the unique characteristics of the ABR mode profile can be advantageous for various applications such as sensing, cavity QED, and telecommunication.

### A. Sensing

Of special interest is the possibility to confine the light in a defect consisting of low refractive index medium (such as air), which can be used to attain strong interaction between the cavity photons and any desired material. This characteristic can be used to realize sensitive and compact sensors, which are able to detect small quantities and low concentrations of analyte.

Among the most straightforward approaches for optoelectronic (bio)chemical sensing is to detect the change in the refractive index or the absorption caused by the presence of a (bio)chemical agent. Several schemes have been suggested to detect these types of changes, employing directional couplers [30], Mach-Zehnder interferometers (MZI) [31] or high-$Q$ optical resonators [7]. The detection mechanism underlying these sensors is the modification of the phase accumulation rate (i.e., the propagation coefficient) of the field due to the interaction of the evanescent tail of the field with the analyte.

The primary disadvantage of these detection methods is that the interaction of the field with the surrounding environment is weak, and therefore, the influence of the analyte on the propagation coefficient is small. As a result, achieving high sensitivity requires large interaction length leading to the requirement of long interferometers and very high-$Q$ resonators. In addition, the MZI-type sensors might have difficulties detecting small numbers (or single) molecules, regardless of their length.

On the other hand, the ABR structure and mode profile (see Fig. 2) allow for the interaction of the *non-evanescent* part of the field, especially when the device is designed to include an air defect. As a result, ABR-based sensors are expected to offer significantly enhanced sensitivity compared to sensors that are based on conventional resonators of similar dimensions and materials.

Figure 4 shows a comparison between the shifts of the resonance frequency of an ABR and a conventional ring resonator due to changes in the refractive index of the surroundings. The ABR consists of alternating layers with refractive indices of 1.545 and 1.0 and an air defect. The conventional resonator consists of $n$=1.545 core surrounded by air cladding. Both resonators are approximately 16µm in diameter and designed to resonate in the visible wavelength regime. The sensitivity of each device is indicated by the slope of the curves shown in Fig. 4. The resonance wavelength of the conventional ring resonator shifts by approximately 0.007 nm for an increase of $10^{-3}$ in the refractive index. For the same index change, the Bragg resonator's resonance wavelength shifts by 0.4 nm, i.e., the ABR exhibits higher sensitivity by a factor of 60.

### B. Telecommunication

Properties such as large FSR and high $Q$ are essential for any resonator-based telecom application, especially filters, add/drop multiplexers, and optical delay lines [1-6], as well as for low threshold lasers.

One of the interesting differences between ABRs and conventional resonators is the in-plane coupling mechanism to other devices. While in a conventional resonator the coupling is evanescent, the coupling between ABRs is direct (i.e., radiative) – similar to the coupling between PC waveguides and defect cavities [32-35]. The radiative coupling mechanism has some advantages and drawbacks compared to evanescent coupling. The main advantage is that the coupling can be determined precisely according to the number of Bragg reflection layers. The drawback is that the resonator cannot be directly coupled to an *in-plane* conventional waveguide (i.e., TIR based), but only to a waveguide also possessing a suitably designed periodic structure, such as a transverse Bragg resonance waveguide [36].

Nevertheless, direct coupling to a conventional waveguide is possible by employing a vertical coupling scheme. Moreover, direct in-plane coupling to other ABRs is possible, thus allowing for structures comprising Bragg-reflection based elements in one layer and conventional I/O waveguides in another (see Fig. 5). Such a configuration is useful for the realization of devices that require precise coupling such as coupled-resonator-optical-waveguide (CROW) delay-lines and lasers [36,37].

## IV. FABRICATION

To examine the spectral and spatial properties of the optical modes supported by the ABR structure, we employed high index-contrast radial Bragg gratings fabricated in active semiconductor material. The semiconductor medium consists of a 250 nm thick InGaAsP layer ($n \approx 3.35$ at $\lambda = 1.55$ μm) on top of an InP substrate. The InGaAsP layer includes six 75 Å wide compressively strained InGaAsP quantum wells positioned at the center, with peak photoluminescence occurring at 1559nm.

The fabrication process is illustrated in Fig. 6. First, a SiO$_2$ etch mask layer is deposited by PECVD (a). Then, a layer of PMMA electron beam resist is applied by spin-coating (b). The desired geometry is then defined using a direct electron beam writer operating at 100 kV (c). After the resist is developed, the PMMA patterns are transferred into the SiO$_2$ etch mask layer by inductively coupled plasma reactive ion etching (ICP-RIE) using C$_4$F$_8$ plasma (d). The remaining PMMA is removed with a gentle isotropic O$_2$ plasma step. The SiO$_2$ layer serves as a hard mask for pattern transfer into the active InGaAsP layer, using an ICP-RIE etch employing HI/Ar chemistry [39] (e). The patterns are etched to a depth of ~ 325nm, completely penetrating the active membrane. The remaining SiO$_2$ hard mask is then stripped in a buffered hydrofluoric acid solution.

To achieve strong vertical confinement, the InGaAsP membrane must be clad by low-index material both above and below. An epitaxial layer transfer technique [40], using a UV-curable optical adhesive (Norland Products NOA 73), is used to flip-bond the patterned semiconductor sample to a transparent sapphire substrate (f). Subsequently, the InP substrate is removed by mechanical polishing and selective wet chemical etching, leaving the 250nm thick patterned

InGaAsP membrane embedded in the cured adhesive (g). Finally, the adhesive filling the trenches is removed with an isotropic NF$_3$/O$_2$ ICP-RIE etch (h). Fig 7 depicts scanning electron microscope (SEM) images of an ABR device at various stages of the fabrication.

Since the optical emission and gain from the compressively strained quantum wells favor TE-polarized electric fields [41], the design of the fabricated devices is optimized for this polarization. In order to simplify the design calculations, we employ the effective index approximation in the vertical dimension. An effective index $n_{eff} = 2.8$ is found by solving for the TE-polarized mode of the transferred InGaAsP slab. To facilitate the fabrication of the device, a mixed Bragg order scheme is used, with second-order ($3\lambda/4 \sim 430$nm) high-index layers and first-order ($\lambda/4 \sim 400$nm) low-index layers.

In addition to relaxing the fabrication tolerances, the mixed Bragg order implementation induces a coherent diffraction component in the vertical direction [15]. Although this mechanism reduces the overall $Q$ of the cavity, it facilitates the observation and measurement of the resonator emission.

## V. EXPERIMENTS

The near-field (NF) intensity pattern and the emitted spectrum of the ABRs are examined at room temperature under pulsed optical pumping. Figure 8 depicts the experimental setup used to characterize the fabricated devices. The pump beam is focused on the sample with a 50X objective lens. The position of this lens is used to control the size and the position of the pump spot. A 20X objective lens is used to collect the vertical emission from the sample and to focus it on an IR camera to obtain the NF intensity pattern and to couple the light into a multi-mode fiber to obtain the emitted spectrum.

The resonators are pumped by pulsed optical excitation, using a mode-locked Ti:sapphire laser emitting ~ 120fs FWHM pulses at a repetition rate of 76.6MHz, with a center wavelength of $\lambda_p$=890nm. The pump beam incidents normal to the plane of the devices under test. When the un-patterned QW layer structure is pumped, the emitted spectrum consists of a wide peak centered at 1559nm. As the pumping power is increased from 1mw to 20mW, the FWHM of the luminescence broadens from approximately 70nm to 110nm, and the peak of the photoluminescence shifts towards longer wavelength due to heating. No significant shift is observed when the pump power is below 5mW, indicating that heating is of less significance at these pump levels.

When an ABR is pumped, the emission characteristics change significantly. While the specific details (threshold levels, emitted wavelengths, etc.) vary from device to device, the overall behavior is similar. Once a certain pump intensity threshold is exceeded, clear and narrow (~0.5nm FWHM) emission lines appear in the spectrum (see Fig. 9). As the pump intensity is increased, the intensity of the emission lines increase as well and they broaden towards shorter wavelengths. Increasing the pump power further results in the appearance of additional emission lines.

Figure 9 shows the lasing characteristics at different pump levels of an ABR consisting of 5 internal and 10 external Bragg layers and a half-wavelength wide defect. The radius of the defect is approximately 5μm. At low pump levels, below 0.75mW, only a single emission line at 1.595μm is visible (the device was design to have a resonance wavelength at 1.6μm). As the pump level is increased, additional resonances at both lower and shorter wavelength appear although the peak at 1.595μm remains the dominant one. The inset of Fig. 9 depicts the integrated emitted power from the lasers vs. the pump level, indicating a clear lasing threshold at $P_{pump}$=680μW. Although the laser was designed for a specific mode Fig. 9 indicates the existence of additional lasing modes. The existence of these modes stem from the combination of the large index contrast between the Bragg layers, which generated an effective radial "bandgap", and the use of three-quarter-wavelength layers consisting of high-index material, which effectively elongate the defect circumference. As a result, the device also supports additional radial and azimuthal modes. However, because the radial index profile is optimized to a specific modal field profile, these additional modes are lossier and thus their threshold pump levels are higher.

We also studied the impact of the pump-spot dimensions on the emission characteristics. By changing the size of the pumped area it is possible to selectively excite the resonant modes of the cavity according to their radial profile. Figure 10 shows measured spectra from the same device as that of Fig. 9, for increasing pump-spot diameters. The pump level is maintained constant at 1.2mW. Thus, as the pump spot is increased the pump density decreases at the center of the device and increases in the periphery, effectively scanning over the resonator area. As the pump-spot is broadened, spectral features having longer wavelength and smaller FSR appear, with the most prominent transition occurring between 10μm and 11μm diameter pump-spots.

We assume that resonances that appear at larger pump-spots peak at larger radii within the device. Under this assumption, the resonance frequencies of the device can be categorized into three distinct groups according to their radial profile. These groups, marked as "S", "M" and "L", are located at small, medium and large radii respectively (see Fig. 10). The insets of Fig. 10 show the IR emission pattern from the ABR at pump-spot diameters of 9.6μm and 11.1μm. The emitted pattern at $D_{pump}$=9.6μm consists of two bright rings: an inner ring with an angular propagation coefficient of $m$=3, and an outer ring whose angular propagation coefficient cannot be resolved. The outer ring is located at the radial defect of the device. We attribute these modes to the strongest peaks in the corresponding spectrum at λ=1595nm and at λ=1615nm. The pattern at $D_{pump}$=11.1μm includes the inner ring with $m$=3 but does not exhibit the outer ring observed for the smaller pump-spot. Since wider pump-spots are associated with longer wavelength we infer that the inner ring corresponds to λ=1615nm and that the defect mode corresponds to λ=1595nm. As can be expected, the pattern at $D_{pump}$=11.1μm is wider than the one at $D_{pump}$=9.6μm, and exhibits modes which are located at larger radii.

## VI. Discussion and summary

We have studied, experimentally and theoretically, the characteristics of a novel class of lasers that are based on radial Bragg reflectors. Lasing action with low threshold levels are demonstrated at room temperature under pulsed optical pumping. The observed $Q$ factors are between 1000 and 2000.

By changing the pump-spot diameter we find that longer resonance wavelengths correspond to patterns with larger radii for the specific structure presented here. For this device, it is possible to correlate between some of the resonance wavelengths and the observed IR patterns, and to identify one of the defect-modes.

Such lasers sources are ideally suited to the detection of small changes in the modal effective index or the $Q$ factor, and for achieving strong atom-field coupling. In addition, our cavity can easily be integrated with other photonic devices such as photonic crystals and distributed feedback lasers.

### Acknowledgment

The authors would like to thank Dr. Axel Scherer and Dr. Oskar Painter for providing access to their fabrication facilities. Fruitful discussions with Joyce Poon and George Paloczi are also acknowledged.

FIGURE CAPTION

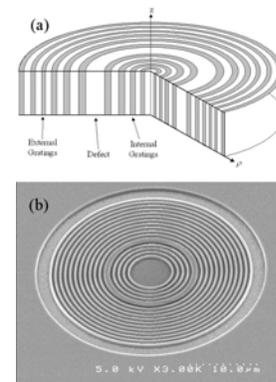

Fig. 1. (a) Schematic of an ABR; (b) A scanning electron microscope (SEM) image of an ABR realized in InGaAsP.

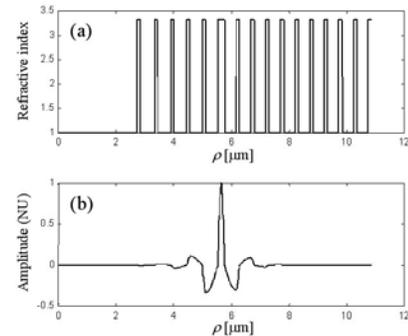

Fig. 2. Refractive index (a) and TE-polarized modal field (b) profiles of an ABR designed for $m=7$, $\lambda_{res}=0.852\mu m$, with 5 internal and 10 external Bragg layers.

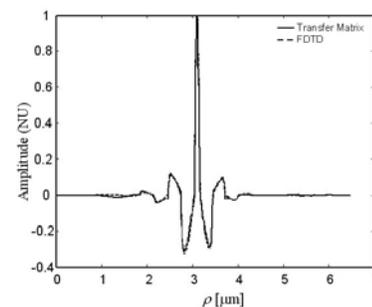

Fig. 3. Comparison between the TE mode profiles calculated by the 2D transfer matrix approach (solid) and the exact solution obtained by 3D FDTD (dashed).

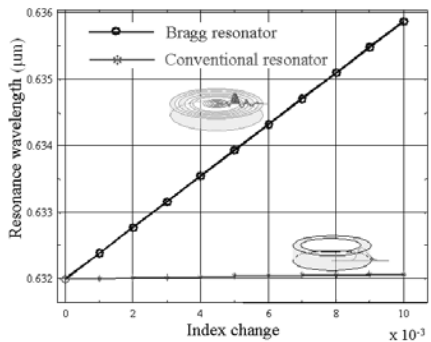

Fig. 4. Comparison of the <u>calculated</u> sensitivity of an ABR and a conventional ring resonator to changes in the refractive index of the surroundings.

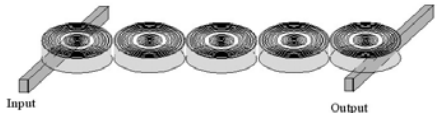

Fig. 5. Illustration of ABR based CROW employing a vertical coupling scheme to conventional waveguides.

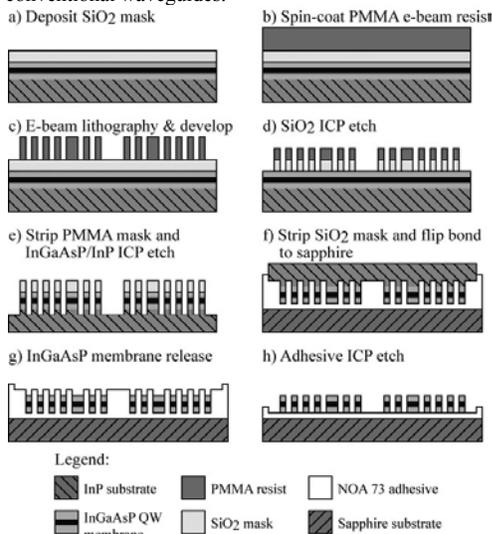

Fig. 6. Fabrication flow diagram and polymer bonding process. The dark regions in the middle of the InGaAsP membrane indicate the QWs.

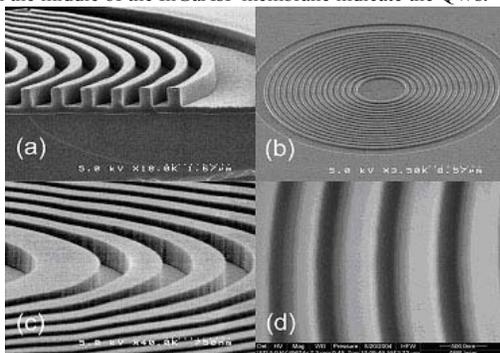

Fig. 7. SEM images of an ABR in various stages of the fabrication process. (a) Cross-section of PMMA pattern after electron beam lithography and development. The SiO$_2$ mask layer can be seen between the PMMA and the substrate. (b) Image taken after the SiO$_2$ mask removal. The radial defect is the 6th ring from the center. (c) Magnified image of etched semiconductor grating, illustrating vertical and smooth sidewalls. (d) Magnified ESEM image of semiconductor rings, taken after the membrane was transferred to the sapphire substrate and optical adhesive was etched.

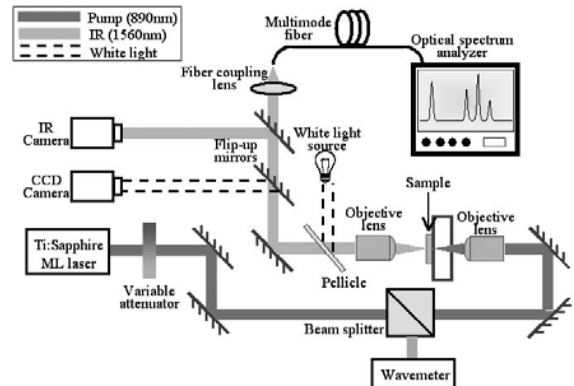

Fig. 8. A schematic of the experimental setup. The dark and the light gray lines indicate the pump and emission beam paths respectively.

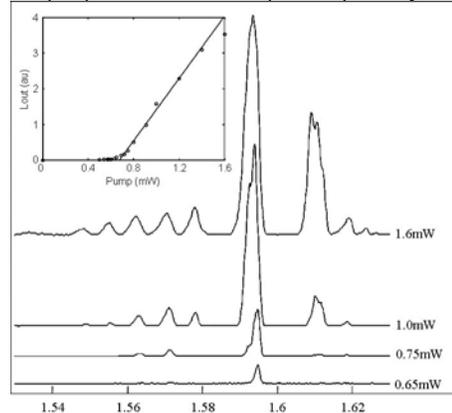

Fig 9. Optical spectra emitted from a lasing ABR under different pump levels. Spectra are vertically offset to illustrate the effects of increasing pump power. Inset: Integrated emitted power vs. pump power, showing laser threshold at ~ 680 µW.

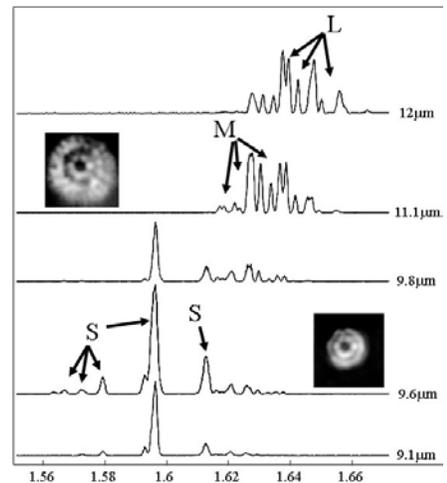

Fig 10. Measured spectra for various pump-spot diameters. The labels "S", "M" and "L" indicate modes located at short, medium and large radii respectively. Insets: IR image of the emitted pattern at 9.6µm and 11.1µm wide pump.